\hsize=13.1truecm %J. Phys. A
\vsize=20.7truecm %J. Phys. A
\baselineskip=12.2pt %J. Phys. A
\parindent=1em
\parskip=0.1ex plus 0.5ex
\font\ttlfnt=cmr10 scaled\magstep 2
%%%%%%%%%%%%%%%
\def\d{{\rm d}}

\def\e{{\rm e}}
\def\eps{\varepsilon}

\def\frac#1#2{{\displaystyle{\displaystyle#1\over\displaystyle#2}}}
\def\p#1#2{\frac{\partial#1}{\partial#2}}
\def\C{{\cal C}}
\def\E{{\cal E}}
\def\A{A}

\noindent{\ttlfnt A simple stochastic model for the dynamics of condensation}
\input epsf

\bigskip
\noindent J.-M.~Drouffe$^{1}$, C.~Godr\`eche$^{2,3}$ and F. Camia$^{1}$
\bigskip
\noindent
$^{1}$ Service de Physique Th\'eorique,
CEA-Saclay, 91191 Gif-sur-Yvette Cedex, France

\noindent
$^{2}$ Service de Physique de l'\'Etat Condens\'e,
CEA-Saclay, 91191 Gif-sur-Yvette Cedex, France

\noindent
$^{3}$ Laboratoire de Physique Th\'eorique et Mod\'elisation,
Universit\'e de Cergy-Pontoise, France

\bigskip\null\bigskip
\noindent{\bf Abstract.}
We consider the dynamics of a model introduced recently by Bialas, Burda and
Johnston. 
At equilibrium the model exhibits a transition between a fluid
and a condensed phase.
For long evolution times the dynamics of condensation possesses
a scaling regime that we study by analytical and numerical means.
We determine the scaling form of the occupation number probabilities.
The behaviour of the two-time correlations of the energy demonstrates that aging
takes place in the condensed phase, while it does not in the fluid phase.

\vfill

\noindent To be submitted for publication to 
Journal of Physics A
\eject

\noindent{\it Introduction. }
 Recently a number of studies have been devoted to the dynamics of the
Backgammon model, a simple stochastic model which exhibits some of the features
of glassy systems such as slow dynamics, non-stationary properties of two-times
correlations, violation of the fluctuation-dissipation ratio, etc. [1-8].
The model introduced in [9] is a simple generalisation of one of the models
defined in [3] (model B), itself closely related to the Backgammon
model. 
In contrast with the latter --or with model B-- it exhibits, at finite
temperature, a phase transition between a fluid and a condensed phase.
The aim of this paper is to study the dynamics of condensation in this model,
hereafter referred to as model B$^\prime$.  
This study is motivated by the fact that,
while for model B the nonequilibrium properties such as slow dynamics [3] or aging
[5] only occur at zero temperature, here they appear in a whole phase, where the
system condenses. In this work we focus our interest on the scaling behaviour of the
occupation probabilities and just attempt a short description of
the two-time correlations. 
The dynamics of this model may also serve as a source of inspiration for
the understanding of the dynamics of the Bose-Einstein condensation, for which little
is known. 
Finally the dynamics of the original branched polymer model introduced by
the authors of ref. [9] may have an interest in its own.

\bigskip

\noindent{\it Definitions. }
Consider a system of $N$ particles distributed amongst $M$ boxes.
We denote by $N_i$ the number of particles contained in box number $i$
($i=1,\ldots, M$), with $\sum_i N_i=N$.
The energy of a given configuration $\C=\{N_1, N_2, \ldots, N_M\}$ 
of the system is defined as the sum of the energies
of individual boxes $\E(\C)=\sum_i E(N_i)$.
The Boltzmann factor associated to $E(k)$ is denoted by
$p_k$. 
The partition function of the system reads\footnote{$^1$}{Note the difference between
the statistics used in eq.~(1) and that used in the definition of the Backgammon
model [1, 7, 8]. 
In contrast to the latter, here the particles are not 
identified by a label. 
In this sense they are indistinguishable [3].}:
\def\eqi{1}
$$
Z_{M,N}=\sum_{N_1}\ldots\sum_{N_M}
\,p_{N_1}\ldots p_{N_M}
\delta\big(\sum N_i,N\big)
=\oint\frac{\d z}{2\pi i\,z^{N+1}}\, \big[P(z)\big]^M
.\eqno(\eqi)
$$
The right hand side expression is obtained by using the integral
representation $\delta(m,n)$ $=\oint\d z\, z^{m-n}/2\pi iz$ for the
constraint.
Equation (1) shows that the equilibrium properties of the system
depend only on the set $\{p_k\}$, or, equivalently, on its
generating function $P(z)=\sum_{k} p_k z^k$.

In the present work we restrict our study to the class of models for which $E(k)$
behaves logarithmically\footnote{$^2$}{Adding to $E(k)$ a linear term in $k$ plays no role because of the constraint.} for large $k$.
 The function $P(z)$ has therefore a finite radius
of convergence
$z_c=1$, with a singularity of the form $(z_c-z)^{\beta-1}$. 
For definiteness we study model B$^\prime$ defined by taking
$E(k)=\ln(1+k)$, hence $p_k=(1+k)^{-\beta}$ leading to a Dirichlet series for $P(z)$
[9].
In parallel we will recall below the properties of model B [3] defined by taking
$E(k)=-\delta_{k,0}$, hence
$p_k=\e^{\beta\delta_{k,0}}$ and
$P(z)=\e^\beta+z/(1-z)$.

\bigskip
\noindent{\it Equilibrium properties. }
The equilibrium properties of the models follow simply from the previous
definitions. 
In the thermodynamic limit $(M,N\to\infty)$, the density $\rho=N/M$
being fixed, the method of steepest descent can be applied to the integral (\eqi).
The saddle-point equation reads $\d P(z)/\d z=\rho P(z)/z$.
The saddle-point value $z_s$,
which is by definition the thermodynamical fugacity of the model,
is thus related to temperature and density. 
The free energy per box reads
$-\beta f=\ln P(z_s)-\rho\ln z_s$.
When $z_s$ increases from 0 to $z_c$, $\d f(z_s)/\d z_s=-\ln(z_s)$
is positive, hence $f(z_s)$ is monotonous. One also finds that in this range,
$\rho$ increases monotonously from 0 to
$\rho_c=z_c P'(z_c)/P(z_c)$. 
While $\rho_c$ is infinite for model B, it is finite in the case of model
B$^\prime$ as long as $\beta>2$ and reads
$\rho_c=\zeta(\beta-1)/\zeta(\beta)-1$ where $\zeta$ is the Riemann
function. 
This fundamental difference between the two models is a consequence of
the behaviour of the $p_k$ at infinity and is at the origin of
the possible existence of condensation in model B$^\prime$.
Indeed when $\rho_c$ is finite, $f(\rho)$ reaches its maximum at $f(\rho_c)=\ln
P(z_c)$. Therefore as long as $\rho<\rho_c$ the system is ``fluid''.
When $\rho>\rho_c$ a condensed phase appears [9].
Thus model B has only a fluid phase for $T>0$.

These two phases are characterized by different forms of the 
occupation probabilities, defined as follows.
The probability that a generic
box, say box number 1, contains $k$ particles is defined as
$f_k={\rm Prob}\big\{N_1=k\big\}$,
i.e., $f_k$ represents
the fraction of boxes containing $k$ particles.
The same definition holds out of equilibrium.
The conservation of the number of boxes and of the number of particles imposes
that
$\sum_k f_k=1$ and
$\sum_k k f_k=\rho$.
From the definition above, one gets
$$
f_k
=\sum_{N_1}\ldots\sum_{N_M}
\,p_{N_1}\ldots p_{N_M}
\,\delta(N_1,k)\,\delta\big(\sum N_i,N\big)
=p_k \frac{Z(N-k,M-1)}{Z(N,M)}.
\eqno(2)
$$
In the thermodynamic limit, using again the steepest descent method, one obtains
\def\eqiii{3}
$$
f_k=p_k\frac{z_s^k}{P(z_s)},\qquad (\rho<\rho_c)
\eqno(\eqiii)
$$
in the fluid phase.
In the condensed phase, one has
\def\eqiv{4}
$$
f_k=\frac{p_k}{P(1)},\qquad (\rho>\rho_c)
\eqno(\eqiv)
$$
which is the same as eq. (\eqiii) with $z_s=1$.
The normalisation condition $\sum_k f_k$ is fulfilled by both equations.
However the conservation of the number of particles, $\sum_k k f_k=\rho$, holds only
in the fluid phase, while it is violated in the condensed phase where this sum is
equal to
$\rho_c$.
The $M(\rho-\rho_c)$ missing particles sit in a single box [9].

An analogous situation occurs for model B at $T=0$.
Eq. (3) gives $f_k=\rho\e^{\beta\delta_{k,0}}(1-z_s)^2z_s^{k-1}$, with
$\rho(1-z_s)=1-f_0$.
When $T\to0$, $\rho_c\to0$, $z_s\to1$, hence $f_0\to1$. 
Again, in order to restore the conservation of particles, all the particles have to
be in a single box.

\bigskip
\noindent{\it Definition of the dynamics. }
The rules defining the dynamics of the models follow naturally from their static
definitions. 
These rules were given for model B in [3].
In this work we used both the Metropolis rule, more convenient for Monte Carlo
simulations and the heat bath rule, leading to simpler dynamical equations.
Let us first describe the former one.
At every time step $\delta t=1/M$ two boxes are
chosen at random, a departure box
$d$, containing $k$ particles, chosen amongst the non-empty boxes, and an arrival
box $a$, containing $l$ particles. 
Note the difference with the Backgammon model, where the departure box is defined by
choosing a particle at random (see footnote 1).
The transfer of one of the particles from box $d$ to box $a$ is
accepted with a probability
$\min\bigg(1,\frac{p_{k-1}}{p_k}\frac{p_{l+1}}{p_l}\bigg)$.

In the heat bath case, once a particle is drawn, it is put into one of the boxes
with a probability proportional to the equilibrium probability of the
resulting configuration.
Thus this move is accepted with a probability
$\frac{p_{l+1}}{p_l}\,\big(\sum_{l=0}^{\infty}f_l{p_{l+1}\over
p_l}\big)^{-1}$. 
The corresponding dynamical equation for the occupation probabilities
is the master equation of a random walk for $N_1$, the number of particles in the
generic box number 1:
\def\eqv{5}
$$
\p {f_k}{t}=\mu_{k+1}f_{k+1}+
\lambda_{k-1}f_{k-1}(1-\delta_{k0})
-\big(\mu_k(1-\delta_{k0})+\lambda_k\big)f_k 
,\eqno(\eqv)
$$
where $\mu_k=1,\, (k>1)$ is the hopping rate to the left, corresponding to
$N_1=k\to N_1=k-1$, and
$$
\lambda_k=
\frac{1-f_0}{\textstyle{\sum_{l=0}^{\infty}f_l{p_{l+1}\over p_l}}}
\frac{p_{k+1}}{p_k},\qquad (k>0)
\eqno(6)
$$
is the hopping rate to the right corresponding to
$N_1=k\to N_1=k+1$.
The factor $1-\delta_{k0}$ accounts for the fact that one cannot select an empty
box as a departure box nor can $N_1$ be negative, i.e. $\lambda_{-1}=\mu_0=0$.
In other terms a partially absorbing barrier is present at site
$k=0$.
This random walk is biased, to the right or to the left according to whether its
velocity $\lambda_k-\mu_k$ is positive or negative, respectively.
It is easy to check that eq. (\eqv) fulfills both conservations of boxes and
particles.

In the stationary state ($\dot f_k=0$) one recovers the equilibrium results
given above.
The detailed balance condition yields
${f_{k+1}}/{f_k}=\lambda_k$,
the two possible solutions of which are precisely those given in eqs. (\eqiii) and
(\eqiv) above, from which one gets 
$$
\frac{1-f_0}{\textstyle{\sum_{l=0}^{\infty}f_l{p_{l+1}\over p_l}}}=
\cases{
z_s & if \quad $\rho<\rho_c$ \cr
\noalign{\vskip1pt} 
1 & if \quad $\rho>\rho_c$ \cr 
}
.\eqno(7)
$$

Model B yields
$$
\lambda_k=\frac{(1-f_0)\e^{-\beta \delta_{k0}}}{1-f_0+f_0\e^{-\beta}}
.\eqno(8)
$$
Whatever the value of $\beta$ the walk is biased to the left.
Therefore, intuitively, no condensation is expected in this model, except at
zero temperature.
In this case, since $\lambda_k=1$ if $k>0$, and
$\lambda_0=0$, the system performs a symmetric random walk with a totally absorbing
barrier at the origin [3].
Hence for $t\to\infty$, $f_0\to 1$, i.e. all boxes become empty.
Nevertheless, in order to fulfill the conservation of particles, one box has
to contain all the particles, as already explained above.
$T=0$ appears thus as a critical point.
Let us point out that for model B, the Metropolis algorithm and the heath bath
rule lead to the same dynamical equation for the $f_k$.

The case of model B$^\prime$ is richer.
A simple analysis, and a numerical check, show that if $\beta$ is large enough,
for small $k$ the bias is to the left, while it is to the right for large
$k$.
Thus one intuitively expects condensation in this model, for a whole range of
values of $\beta$.

\bigskip
\noindent{\it Condensation in the scaling regime. }
A numerical integration of eq. (\eqv), and Monte Carlo simulations for $\beta=4$ and
$\rho=2>\rho_c(4)=.110$, give some insight in the phenomenology of the dynamics in
the condensed phase. Three regimes are observed. 
First a transient one, with a rapid reorganisation of the particles
in the boxes, leading to a situation with a fluid part for small $k$ and the
appearance of a condensate, i.e. a group of boxes containing a large fraction of
the total number of particles. 
This regime is followed by the scaling regime, our main interest in this work, where the
evolution in time of the condensate is self-similar. The number of boxes containing the
condensate decreases (although it remains large in this regime).

\midinsert
\epsfysize=65truemm
$$\epsfbox{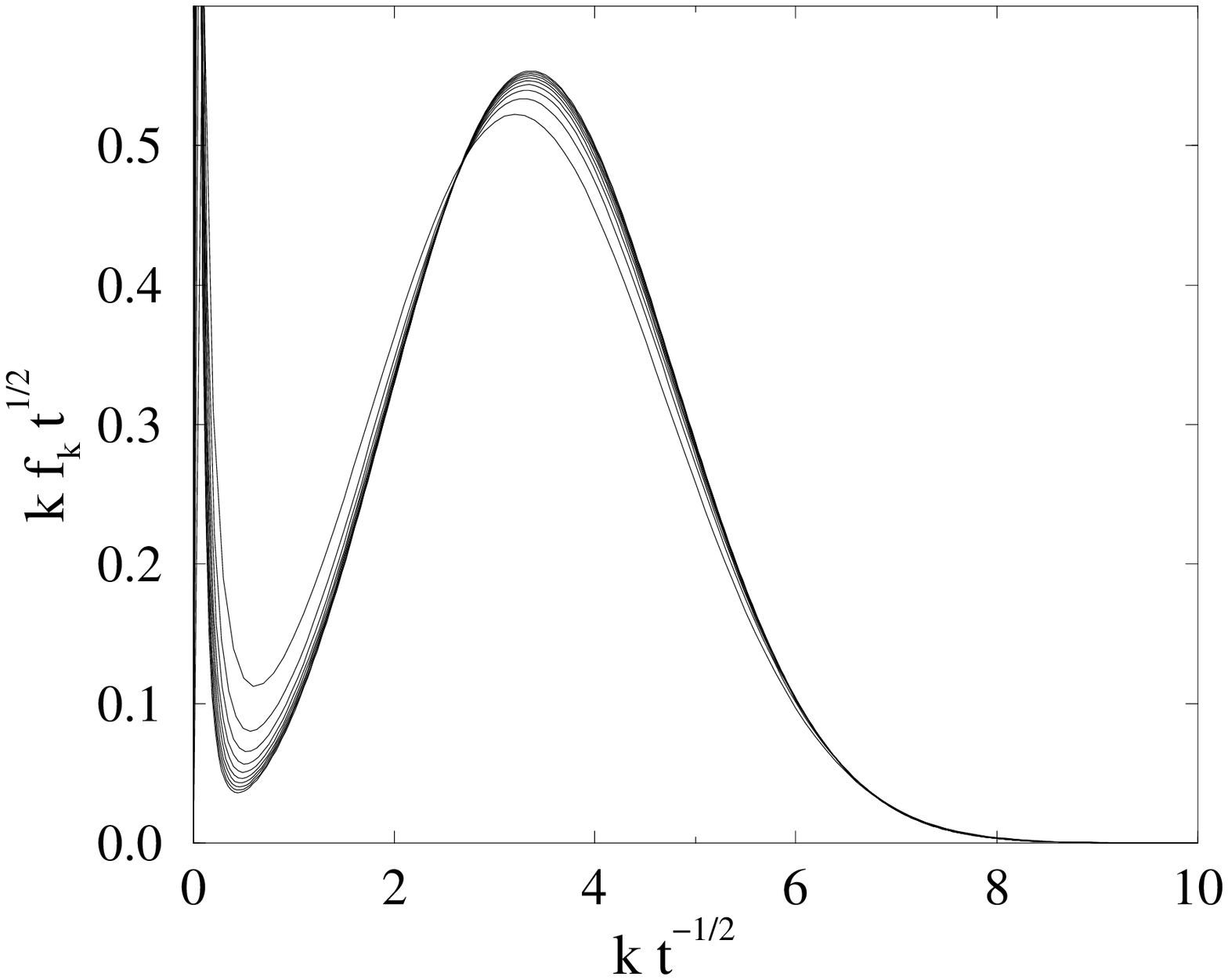}$$
%\vskip-10truemm
{{\bf Figure 1.} Scaling in the dynamics of condensation for model B$^{\prime}$.
$k f_k\sqrt{t}$, obtained by numerical integration of eq. (5), is plotted versus $k/\sqrt{t}$ for 10 different times 
varying from 100 to 1000.
($\beta=4$ and $\rho=2$.)}
\endinsert

Finally at very long times, the condensate reduces to a single box according
to a non universal process where finite size effects
should now be taken into account.
This sequence of three regimes takes place in a similar fashion in the case
of model B.
It is also reminiscent of the dynamics of coarsening systems [10].

In order to describe the scaling regime for $f_k$ we set
\def\eqix{9}
$$
 f_k= 
\cases{
\frac{p_k}{P(1)}(1+\eps v_k+\cdots) &if \quad $k<u_0/\eps$ \cr
\noalign{\vskip2pt} 
\eps^2 g(u)(1+O(\eps))& if \quad $k>u_0/\eps$ \cr 
}
,\eqno(\eqix)
$$
where the small scale $\eps(t)$ is to be determined, and $u=\eps k$ is  the
scaling variable. 
$u_0$ fixes the separation between condensed and fluid phases
and corresponds intuitively to the position of the ``dip'' clearly visible on
figure 1.

The normalisation conditions $\sum f_k=1$ and $\sum k f_k=\rho$
lead respectively to
\def\eqx{10}
$$
\sum_0^{u_0/\eps} \frac{p_k}{P(1)} v_k+\int_{u_0}^\infty g(u)\d u=0
,\eqno(\eqx)
$$
and
\def\eqxi{11}
$$
\rho-\rho_c=\int_{u_0}^\infty u g(u)\d u
.\eqno(\eqxi)
$$
It is easy to check that, once the limit $t\to \infty$ is taken, 
with $u_0$ fixed
(hence $u_0/\eps\to \infty$),
the limit $u_0\to 0$ can then be taken in (\eqx, \eqxi).

In order to get a continuum description of the master equation we use the
expansion, valid when $k$ is large,
${p_{k+1}}/{p_{k}}=1-{\beta }/{k}+{{\rm const.}}/{k^{2}}+\cdots$.
We also set
$$
\frac{1-f_{0}}{\textstyle{\sum_{m=0}^{\infty }f_{m}\frac{p_{m+1}}{p_{m}}}}
=1+\A\eps +\cdots
,\eqno(12)
$$
since this quantity goes to 1 for long times.
The amplitude $\A$ will be determined below.

For small $\eps$ (i.e. large $t$) and fixed $u$ (hence large $k$) one gets 
$$
\frac{\dot\eps}{\eps ^{3}}(2g+ug^{\prime })=g^{\prime \prime}+
\frac{\beta }{u}g^{\prime }-\A g^{\prime }-\frac{\beta g}{u^{2}}+O(\eps )
,\eqno(13)
$$
which is in separable form.
The solution for $\eps $ is const.$\left(t-t_{0}\right) ^{-1/2}$ 
or, after a change in the origin of time,
$\eps =1/\sqrt{t}$.
In this regime one therefore obtains
\def\eqxiv{14}$$
g^{\prime \prime }+\left( \frac{1}{2}u-\A+\frac{\beta }{u}\right) g^{\prime
}+\left( 1-\frac{\beta }{u^{2}}\right) g=0 
,\eqno(\eqxiv)
$$

This equation is singular at $u=0$ and $u=\infty$.
At $u=0$ one expects a power singularity for $g$ of the form $u^{s}$.
The Frobenius series 
$g(u)=\sum_{0}^{\infty }a_{n}u^{s+n}$
carried into (\eqxiv) gives the recursion relation
\def\eqxv{15}
$$
a_{n+2}(s+n+1)(s+n+\beta +2)-\A(s+n+1)a_{n+1}
+\frac{s+n+2}{2}a_{n}=0,\quad (n\ge-2)
\eqno(\eqxv)
$$
with $a_{-1}=a_{-2}=0$.
The radius of convergence of the series thus obtained is infinite.
For $n=-2$ one has $(s-1)(s+\beta)=0$.
Hence, when $u\to 0$, a basis of solutions reads
$$
g_{1}^{(0)}(u)\sim u, \quad
g_{2}^{(0)}(u)\sim u^{-\beta } 
.\eqno(16)
$$
Only the first solution may be retained because of the normalisation of $g$
(cf eqs.  (\eqx-\eqxi)).
When $u\to\infty$, an asymptotic study of the dominating terms in eq. (\eqxiv)
leads to another basis of solutions (reminiscent of the particular case $A=\beta=0$,
for which solutions are $u\e^{-u^{2}/4}$ and $u\e^{-u^{2}/4}\int u^{-2}\e^{u^{2}/4}
\d u$), namely
$$
g_{1}^{(\infty )}(u)\sim u^{1-\beta}\,\e^{-u^2/4+Au}, \quad
g_{2}^{(\infty )}(u)\sim u^{-2}
.\eqno(17)
$$
Again the second behaviour should be excluded, because of the normalisation
conditions (\eqx-\eqxi).
As the two solutions must be connected,
$$
g_{1}^{(0)}(u)=C_{1}(\A,\beta )g_{1}^{(\infty )}(u)+C_{2}(\A,\beta) 
g_{2}^{(\infty)}(u)
,\eqno(18)
$$
and one must impose $C_{2}(\A,\beta )=0$.
This condition determines $\A$ as a function of $\beta$.
In practice, this may be done numerically either by reconstructing $g$ from its series
or directly from a numerical integration of the differential equation (\eqxiv).
The solution of eq. (\eqxiv) may be seen as a continuous deformation of $u\,\e^{-u^2/4}$,
as $A$ and $\beta$ increase from 0.
Note that the solution $g(u)=u\e^{-u^2/4}$ of the equation $g''+u g'/2+g=0$,
corresponding to $\A=\beta=0$, appears also in the scaling regime of model B
at $T=0$, where $f_k=t^{-1}g(k t^{-1/2})$ [3]. Figure 2 displays on the same plot
the solution of eq. (\eqxiv) with $\A$ determined following this technique,
the asymptotic form of (\eqv) for large times and the Monte-Carlo
simulation. 
The agreement between all these curves is excellent.

\midinsert
\epsfysize=65truemm
$$\epsfbox{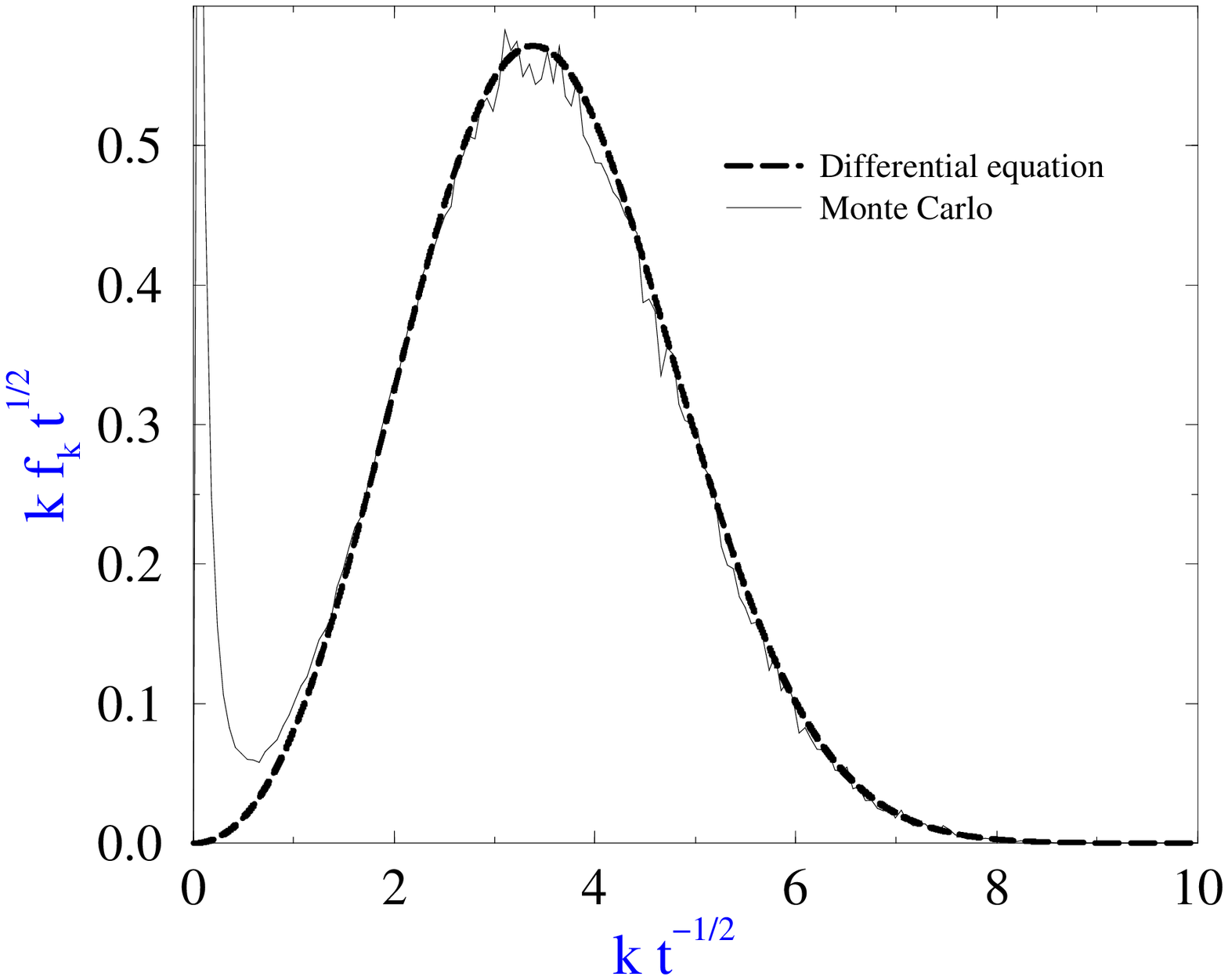}$$
%\vskip-10truemm
{{\bf Figure 2.} Solution of the differential equation (\eqxiv) with $A=1.9$, and Monte-Carlo simulation ($t=1000$, $N=1000, M=500$).
($\rho=2$, $\beta=4$.)
The  numerical solution of (\eqv) for large
times is hardly distinguishable from the solution of (14).}
\endinsert

For theoretical purposes, it may be convenient to cast eq. (\eqxiv)
into its Schwarzian form
\def\eqxix{19}
$$
w^{\prime \prime }+\frac{1}{2}\left( -\frac{u^{2}}{8}+\frac{1}{2}\A u+\left( 
\frac{3}{2}-\frac{\beta }{2}-\frac{\A^{2}}{2}\right) +\frac{\A\beta }{u}-\frac{
\beta (\beta +2)}{2u^{2}}\right) w=0
,\eqno(\eqxix)
$$
obtained by setting $g=v w$ and choosing $v$ such that no first derivative of $w$
appears in the equation.
This leads to
$$
g(u)=u^{-\beta /2}\exp \left( -\frac{u^{2}}{8}+\frac{\A u}{2}\right) w(u)
.\eqno(20)
$$
Eq. (\eqxix) may be recognized as a Schr\"odinger equation with zero energy.
Since $w$ is positive, it corresponds to the ground state solution of the equation. 
Imposing that the energy of the ground state be zero determines $A$ as a function of
$\beta$.

Finally one may find the explicit form of the $f_k$ for small $k$, i.e. in the fluid
part of the distribution. 
Substituting the form (\eqix) for $f_k$ into the
master equation (\eqv) leads to
$$
\eqalign{
v_{1}&=v_{0}+\A\cr
p_{k+1}v_{k+1}+p_k v_{k-1}-(p_{k+1}+p_k) v_k&=A(p_{k+1}-p_k),
\qquad (k>0)
}
\eqno(21)
$$
the solution of which is
$v_{k}=v_{0}+k\A$.
The determination of $v_0$ is made possible by eq. (\eqx) which gives
$$
v_{0}+\A\rho _{c}+\int_{0}^{\infty }g(u)\d u=0
,\eqno(22)
$$
where $A$ is already known from above, showing that $v_0$ is negative.

\bigskip
\noindent{\it Two-time correlations of the energy. }
At $T=0$, the energy correlation function of  model B exhibits aging [5].
It is therefore natural to expect the same property for model B$^{\prime}$ in the
whole low temperature phase, $\rho$ being fixed.
Following the notations of [7], the correlation function of the energy of a generic
box, say box number 1, at two times $s$ and $t$ ($s<t$) reads
$c(t,s)=f_0(s)(g_0(t,s)-f_0(t))$, where 
$g_k(t,s)$ is the probability that box number 1 contains $k$ particles at time $t$,
knowing that it was empty at $s$.
The evolution in time of this quantity is given by (5),
with initial conditions $g_k(s,s)=\delta_{k,0}$.
A numerical integration of these equations shows that the normalised correlation
function $c(t,s)/c,s,s)$ has the
same asymptotic scaling form $\sqrt{s/t}$ as model B [5], for $\beta>\beta_c$,  i.e.
it exhibits aging in the low temperature phase.
We will come back to the theoretical analysis of this
result in a forthcoming publication.

\bigskip
\noindent{\it Discussion. }
We wish to emphasize the remarkable result obtained in this work, namely that the condensate acquires a universal scaling form directly related to the assumption of a regular power law behaviour of the $p_k$ at infinity.
We checked that the master equation with Metropolis rule lead to the same scaling
form for $g(u)$.
As a consequence of this scaling form, the condensation time behaves as the squared size of the system.

\bigskip
\bigskip\noindent
We wish to thank R.~Balian, R.~Conte, J.-M.~Luck and G.~Mahoux for interesting
conversations.

\bigskip
\bigskip\noindent
{\parindent 0em
{\bf References}
\vskip 12pt plus 2pt

[1] F.~Ritort, Phys. Rev. Lett. {\bf 75} (1995), 1190.

[2] S.~Franz and F.~Ritort, Europhys. Lett. {\bf 31} (1995), 507.

[3] C.~Godr\`eche, J.P.~Bouchaud, and M.~M\'ezard, J. Phys. A {\bf 28}
(1995), L603.

[4] S.~Franz and F.~Ritort, J. Stat. Phys. {\bf 85} (1996), 131.

[5] C.~Godr\`eche and J.M.~Luck, J. Phys. A {\bf 29} (1996), 1915.

[6] S.~Franz and F.~Ritort, J. Phys. A {\bf 30} (1997), L359.

[7] C.~Godr\`eche and J.M.~Luck, cond-mat/9707052, to appear in J. Phys. A.

[8] B.J.~Kim, G.S.~Jeon, and M.Y.~Choi, Phys. Rev. Lett. {\bf 76} (1996),
4648.

[9] P.~Bialas, Z.~Burda and D.~Johnston, cond-mat/9609264.

[10] B.~Derrida, C.~Godr\`eche and I.~Yekutieli, Phys. Rev. {\bf A44} (1991), 6241.

}
\bye